**OFFSHORE-WIND ENERGY | RENEWABLE ENERGY**

# Wind energy potential of the German Bight
## Limits and consequences of large-scale offshore wind energy use

Axel Kleidon

*The wind blows stronger and more reliably over the sea than over land. Thus, offshore wind energy is expected to make a major contribution to the energy transition in Germany, especially in the German Bight. But what happens when a growing number of wind farms extract more and more wind energy from the atmosphere?*

The challenges of the energy transition for the next decades in Germany are enormous. It is true that 15.9 % of primary energy demand was already covered by renewable energy in 2021 [1], and a lower energy demand is expected in the future due to more modern technologies such as heat pumps and electromobility. However, the transition to a complete, sustainable energy system that is free of fossil fuels is still a long way off.

Many energy transition scenarios focus on the expansion of a combination of solar and wind energy. These two types of renewable energy have the greatest potential in Germany [2] and complement each other very well over the course of the year: while the Sun can supply a particularly large amount of renewable energy in summer, it fails in winter. This can be compensated for by wind energy, as the dark winter months are usually stormier than the summer.

Wind power generation at sea plays a special role in these scenarios. Wind blows stronger and more continuously at sea than on land, so it can generate electricity more efficiently and reliably. In Germany, expansion is planned mainly in the German Bight of the North Sea, where the exclusive economic zone - i.e. the part of the sea that is administered by Germany beyond the territorial sea - offers considerably more surface area than the Baltic Sea. For example, wind farms with 6.7 GW of installed capacity are currently located in the North Sea, compared to only 1.1 GW in the Baltic Sea (as of 2021, [3]). In 2021, these wind farms contributed about 24 TWh/a or 4.9 % to the German electricity demand of 491 TWh/a, which means that the turbines were utilized to an average of 35 % - the so-called capacity factor [3]. Wind turbines at sea were thus almost twice as productive as on land, where the capacity factor was only 18 %.

By 2050, it is assumed that the use of offshore wind energy will increase significantly more than on land, i.e. onshore. In its coalition agreement, the German government has targeted the expansion of offshore wind energy to 70 GW, i.e. roughly a tenfold increase in currently installed capacity. Onshore, there is already 56 GW of turbine capacity, and an expansion to around 200 GW is expected here, distributed over 2% of the country's surface area. However, with 357,000 km$^2$ there is considerably more space than in the exclusive economic zone of the North Sea, which is only 28,600 km$^2$ in size. So the plans envisage a much more intensive use of wind energy at sea than on land. And because each wind turbine draws energy from the atmosphere and thus weakens the winds, the question arises whether, with such a strong expansion, the turbines could take the wind away from each other and thus endanger the high yields.

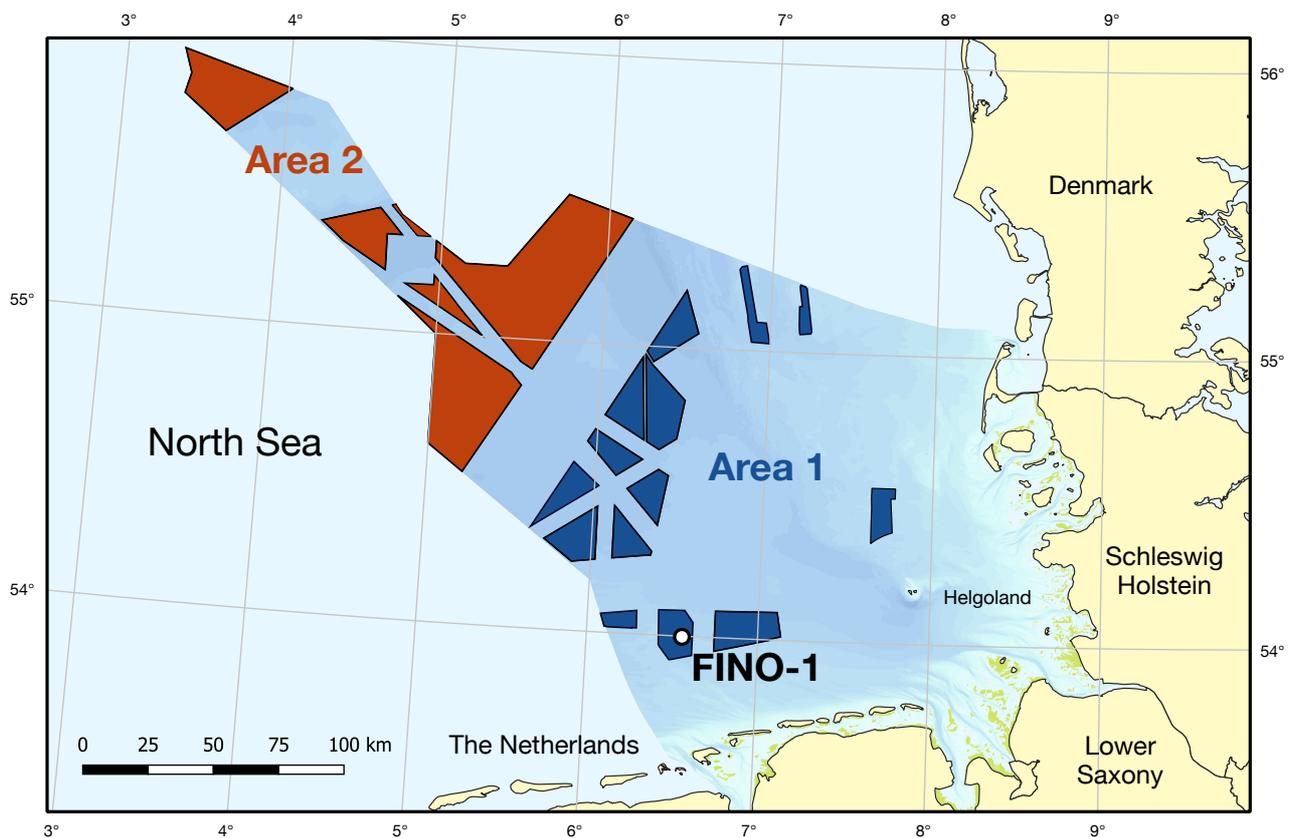

*Fig.1 Areas of the German Bight that can be used for the development of wind energy. The areas close to the coast in dark blue are referred to here as area 1, the red areas far from the coast as area 2. Black and white circle: position of the FINO-1 measuring station (map modified according to [5]).*

## Wind energy in the German Bight

This question was examined in a report by Agora Energiewende on the wind energy potential of the North Sea [5]. I worked scientifically on this report and want to present the results here in a comprehensible way. This study has also already been taken into account in the current, official planning of offshore wind energy in Germany. In the following, I will go through the steps necessary to determine the potential for electricity generation by wind energy use in the German Bight. In particular, I want to make the effect of wind extraction by the turbines physically plausible.

In the first step, we determined the areas that are potentially available for the expansion of wind energy (Figure 1). There is a whole range of different uses of the sea. These include, of course, shipping, which needs routes, certain areas are designated as nature reserves, there are areas used for military purposes, and areas are needed for submarine cables and supply lines. These areas preclude wind energy use, which significantly reduces the total area available. The usable areas can be roughly divided into two areas separated by a wide route for shipping: the coastal area 1 (blue in Figure 1) with 2767 km$^2$ and the far-from-the-coast area 2 (red) with 4473 km$^2$.

Next, we need technical information on the turbines that will be placed in these areas. For this purpose, we choose a hypothetical 12 MW turbine with a rotor diameter of 200 m, which corresponds to the specifications of the currently most powerful turbines. The power generation of a single turbine is described by the so-called power curve. It shows how much electricity an

## FIG. 2 WIND IN THE GERMAN BIGHT

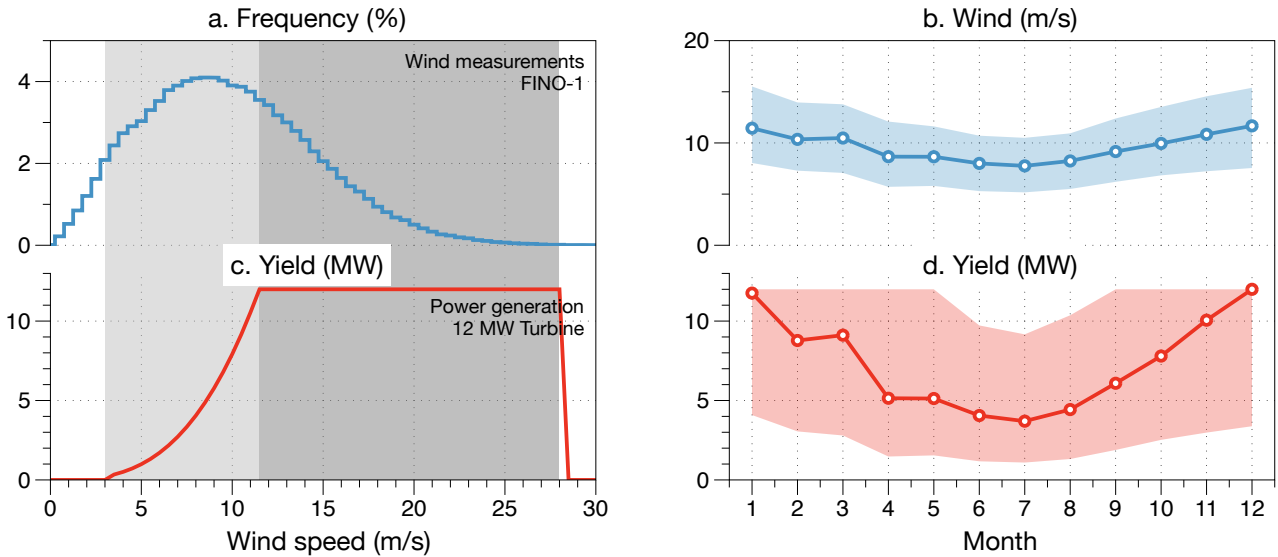

*Wind conditions in the German Bight and their use by an isolated standing wind turbine. a) Frequency distribution shows wind measurements 2004-2015 at 100 m height on FINO-1 in the North Sea [4], position of this measuring station see Figure 1. b) The seasonal course of wind speeds over the months is shown by the median, where the area highlighted in blue covers 25-75 % of the distribution. c) Yield of an isolated 12 MW wind turbine as a function of wind speed and d) its seasonal variation in the North Sea. Highlighted in light grey on the left is the range where wind yield increases with wind speed (61.5% of the time), while in the range highlighted in dark grey the turbine operates at its capacity limit (32.6% of the time).*

isolated turbine produces at a prevailing wind speed, the so-called wind yield (Figure 2c). The dependence on the wind speed can be roughly divided into four ranges: In calm conditions below the cut-in velocity of 3 m/s, the turbine produces no electricity. In the second range up to the rated velocity of 11.5 m/s, the output increases proportionally to the kinetic energy flux density given by $(1/2)\, \rho\, v^3$, with the air density $\rho = 1.2$ kg/m³ and the wind speed $v$ in m/s. The energy flux density is then multiplied by the cross-sectional area spanned by the rotor and the power coefficient of about 0.42 (that is, 42% of the kinetic energy flux density can be used) to determine the yield in this range. In the third range above the rated wind speed up to the cut-out velocity of 28 m/s, the yield is determined by the capacity of the generator. Above this wind speed, the turbine is shut down to protect against damage and does not generate any electricity.

The average yield of the wind turbine is then determined by combining the power curve with the frequency distribution of wind speeds. For this purpose, we used the frequency distribution of wind speeds (Figure 2a) measured by the FINO-1 measuring station in the German Bight at a height of 100 m in the period 2004-2015. Its position is marked in Figure 1. These data show that the absence of wind is relatively rare with an average of 5.8 %, the second range where the yield depends directly on the wind speed is the most frequent with 61.5 %, 32.6 % of the time the turbine operates at its capacity. In only 0.1 % of the time the turbine has to be shut down due to excessive wind speeds.

In total, the turbine generates an average of 6.8 MW of electrical power or 59.1 GWh of electrical energy per year. The efficiency of the energy generation can be described on the one hand by the full load hours, with which the annual yield is simply described by the product with the capacity of the turbine. The annual yield is thus calculated as 12 MW * $x$ h/a = 59.1 GWh/a with $x$ = 4928 full load hours per year. On the other hand, the efficiency can be described by the capacity factor, which describes the ratio of the average yield to the capacity of the turbine. In our case, the capacity factor is 6.8 MW/12 MW = 56.7 %. The efficiency - or the capacity factor - is not only described by the technical specification of the turbine, but also by the wind conditions. For example, the capacity factor in Germany on land is only about 20 % [6]. In principle, the yield is also subject to seasonal fluctuations, with higher yields in winter than in summer (Figure 2d).

Next, we considered different scenarios in which the two areas 1 and 2 are equipped with different numbers of wind turbines. Three scenarios rely solely on the use of Area 1 for wind energy because of its proximity to the coast makes the costs of installation, supply, and connection to the power grid less expensive. These scenarios consider different installation densities of 5, 10 and 20 MW per square kilometre. With an area of 2767 km$^2$, this corresponds to 1153, 2306, and 4612 turbines with 12 MW capacity each.

In five other scenarios, we consider both areas with installation densities of 5, 7.5, 10, 12.5, and 20 MW/km$^2$, with 3017 to 12067 turbines distributed evenly over the 7240 km$^2$ of both areas combined. This gives us a total of eight scenarios, covering a range of 14 to 145 GW of installed capacity. The German government's expansion target of 70 GW is thus well covered.

## Wind yield estimation

Next, we determined the total yield of the installed turbines for the different scenarios. A seemingly obvious way to do so would be to simply multiply the yield of the isolated turbine by the number of turbines. This gives us theoretical results for yields as shown by the light bars in Figure 3. This type of estimation is currently widely used. Sometimes it is reduced by an empirically determined park loss factor of 10 %, but sometimes it is even expected that technological progress will actually increase turbine efficiency. The scenarios then result in a wind yield of 7.8 to 82.1 GW or 68.2 to 713.6 TWh/a. By comparison, electricity consumption in Germany in 2021 was around 491 TWh/a [3].

However, this way of calculating yields does not take into account that wind turbines extract a considerable amount of kinetic energy from the atmosphere. This weakens the wind and thus the average efficiency of the turbines in the region. We can easily see this by looking at the kinetic energy fluxes of the region (Box "KEBA: Kinetic Energy Balance of the Atmosphere" on p. 7). On the one hand, there are the two inputs into the lower atmosphere of the region, the so-called boundary layer, which over the North Sea is about 700 m thick: The first contribution comes from the horizontal flow into the region, the second comes from above through vertical mixing.

Area 1 in Figure 1 has an area of 2767 km$^2$. We consider it simplified as a square in the following, with a length of about 52.6 km. If we assume a wind speed of 9.4 m/s, which corresponds to the median of the frequency distribution in Figure 2a, this is in the range where the wind yield increases with wind speed (Figure 2c). Thus, about 52.6 x 10$^3$ m x 7 x 10$^2$ m x (0.5 x 1.2) kg/m$^3$ x (9.4 m/s)$^3$ ≈ 18.3 GW flows into the area, while the vertical replenishment is relatively small at about 2.8 GW (see equations (2) and (3) in the Box "KEBA: Kinetic Energy Balance of the Atmosphere"). Thus, 21.1 GW of kinetic energy enters Area 1 at this wind speed, which is already quite close to the installed capacity of 14 GW for the smallest scenario for Area 1 with 1153 turbines. So we can see that the wind turbines will extract an appreciable amount of kinetic energy from the region and their effect must be taken into account.

For estimating the yields of different scenarios, we can take the balance of kinetic energy fluxes in our virtual box (box "KEBA: Kinetic energy balance of the atmosphere" [7], and Figure 4). The

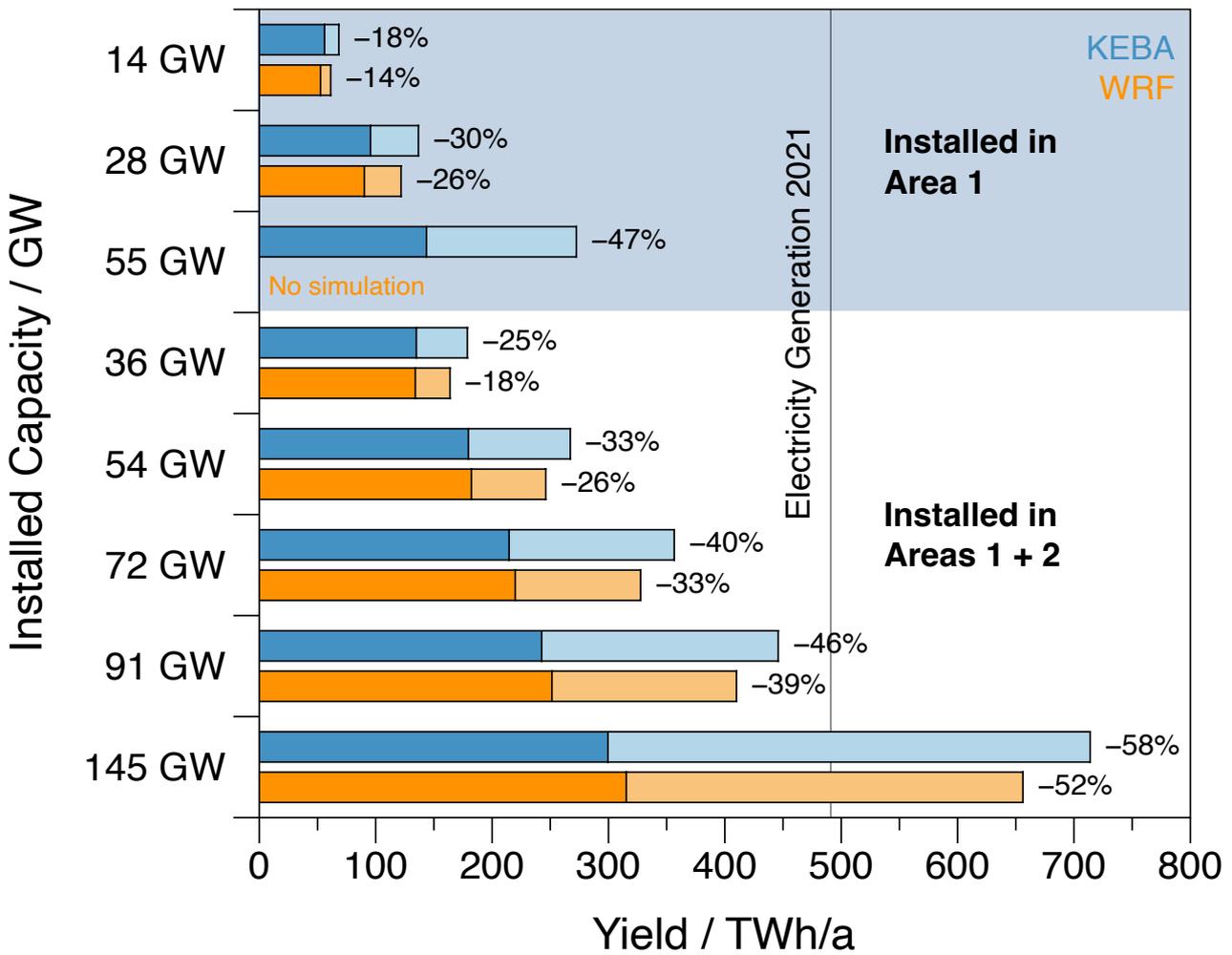

**FIG. 3 ELECTRICITY YIELD OF DIFFERENT SCENARIOS**

*Electricity yield of different offshore wind energy expansion scenarios in the German Bight without (light) and with (dark) the extraction of wind energy by the turbines. The blue estimates are based on the KEBA approach (see "KEBA: The Kinetic Energy Balance of the Atmosphere"), while the orange estimates are based on calculations with a much more complex numerical weather prediction model (WRF). The vertical black line represents Germany's average electricity consumption in 2021. As a comparison: the German government's expansion target for offshore wind energy for 2030 is 30 GW of installed capacity, in 2050 it is 70 GW (data from [5]).*

estimates from this approach are shown by the blue bars in Figure 3. The orange bars come from calculations using a much more complex numerical weather prediction model. As we can see, the results from both methods are very similar. So looking at the energy fluxes in the atmosphere is the key to understanding the reduced yields from strong wind energy use.

For a complete balance of the kinetic energy flows, we also need to look at the loss terms. In addition to the extraction of energy by the turbines, there is also the friction loss in the wake of the turbines, surface friction as well as the export of kinetic energy into the areas downwind of the wind farms. The effect of wind extraction can be represented comparatively simply with a reduction factor, since all these components depend on the kinetic energy flux density.

*FIG. 4 KINETIC ENERGY BALANCE OF THE ATMOSPHERE*

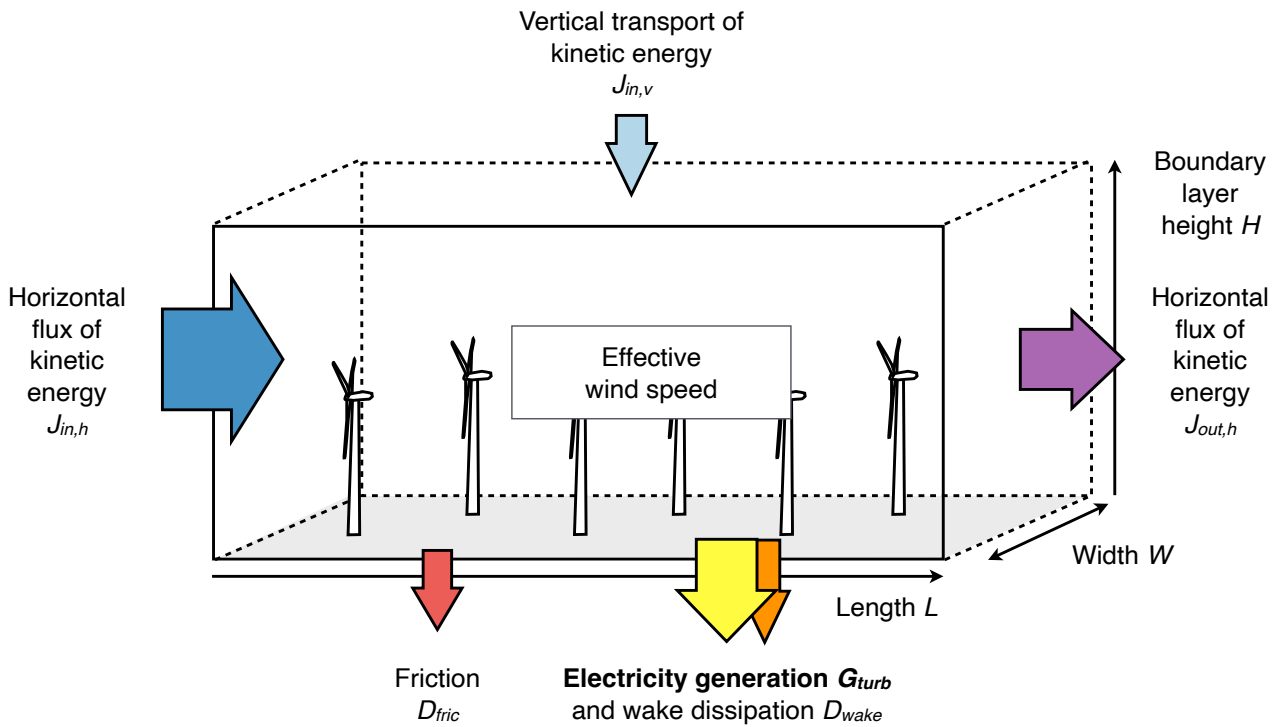

*Kinetic energy balance of the atmosphere over the region in which wind energy is used and from which an effective wind speed can be calculated. See box "KEBA: Kinetic energy balance of the atmosphere" for explanation of symbols and how this calculation is done.*

This factor depends primarily on the size of our virtual box and the number of turbines (see formula (10) in the box "KEBA: Kinetic energy balance of the atmosphere"). It reduces the yield especially at low wind speeds, since it then depends strongly on wind speed. At high wind speeds, much more kinetic energy enters into our box, since the kinetic energy fluxes depend on the third power of the wind speed. In this case, the turbines operate at their capacity, which means that lowering the wind speed does not affect their yield as much.

Figure 5 shows an example of the change in the various contributions in the kinetic energy balance for the scenarios. The natural case without wind energy use is also included. Here, the input of kinetic energy is balanced with surface friction and downwind export. The more wind energy is used in the areas, the more the terms shift towards electricity generation (yellow in Figure 5) and frictional losses to the wakes (orange). This is at the expense of surface friction and export. These two terms are directly coupled to wind speed, so wind speed must decrease. This can clearly be seen in the frequency distribution of wind speeds (Figure 6), which shifts towards lower values with greater use.

# KEBA: Kinetic Energy Balance of the Atmosphere

The effect of wind extraction by turbines can be described quite simply and physically with the help of the kinetic energy balance of the lower atmosphere [7]. For this purpose, we consider the air volume above the area of the planned wind farms (Figure 4), with width $W$ and length $L$, as well as a height $H$. This height comprises the boundary layer in which the lower atmosphere is well mixed. Over the North Sea this is usually around 700 m high.

We now consider the components that contribute, export, or convert kinetic energy in this volume. These are the kinetic energy inputs from upwind areas and from above, $J_{in,h}$ (dark blue arrow in the figure) and $J_{in,v}$ (light blue arrow), the export $J_{out,h}$ downwind (purple arrow), the frictional loss due to surface friction $D_{fric}$ (red arrow), the extraction by the turbines for power generation $G_{turb}$ (yellow arrow), and the frictional losses in the wakes due to the mixing of surrounding air masses $D_{wake}$ (orange arrow).

$$J_{in,h} + J_{in,v} = J_{out,h} + D_{fric} + G_{turb} + D_{wake} . \quad (1)$$

The horizontal input of kinetic energy is described by:

$$J_{in,h} = [(\rho/2)\, v_{in}^3\,] * W * H. \quad (2)$$

The expression in the brackets describes the kinetic energy flux density, with the air density $\rho$ of about 1.2 kg/m$^3$ near the sea surface and the wind speed $v_{in}$. We can describe the input of kinetic energy from the free atmosphere, which is above the boundary layer, by vertical mixing through the friction loss at the surface when there are no wind turbines, because then these two terms balance. This is described by:

$$J_{in,v} = \rho\, C_d\, v_{in}^3 \times W \times L \quad (3)$$

where $C_d$ represents the drag coefficient and which is typically about 0.001 over sea.

If wind turbines are present, we describe the wind speed within the volume by an effective speed $v$. It will be lower than $v_{in}$, because the wind turbines change the kinetic energy balance of the volume. We use this effective wind speed to describe the other four terms of the balance. We write the export of kinetic energy into downwind areas analogous to (2) as

$$J_{out,h} = [(\rho/2)\, v^3\,] \times W \times H. \quad (4)$$

For the friction loss we write similar to (3):

$$D_{fric} = \rho\, C_d\, v^3 \times W \times L. \quad (5)$$

The power generation, or yield, of the wind turbines in the range when power depends on the wind speed, is given by

$$G_{turb} = [(\rho/2)\, v^3\,] \times \eta \times A_{rotor} \times N, \quad (6)$$

where $\eta$ is the power coefficient of the turbine, typically $\eta$ 0.42, $A_{rotor}$ is the cross-sectional area spanned by the rotor blades, in the case of our 12 MW turbine this is 31415 m$^2$, and $N$ is the number of wind turbines.

For the friction loss in the turbine wakes, we assume 50% of the power extracted from the wind by the turbines as a realistic value. Thus it follows:

$$D_{wake} = 0{,}5 \times G_{turb} . \quad (7)$$

The four terms of the right-hand side of the kinetic energy balance (1) all depend on $v^3$, so we can easily obtain the effective wind speed by rearranging the equation. This can then be described as

$$v = f_{red}^{1/3}\, v_{in}, \quad (8)$$

and the amount of electricity generated by

$$G_{turb} = f_{red}\, [(\rho/2)\, v_{in}^3\,] \times \eta \times A_{rotor} \times N. \quad (9)$$

Here, $f_{red}$ is a reduction factor describing the effect of wind extraction from the volume:

$$f_{red} = \frac{H + 2 C_d \cdot L}{H + 2 C_d \cdot L + 3/2 \cdot \eta A_{rotor} \cdot (N-1)/W}. \quad (10)$$

Note that for an isolated turbine ($N = 1$) this factor is 1, so there is no yield reduction. The higher the number of turbines and the larger the rotor area, the smaller the factor becomes, the wind is weakened and the yield is reduced. In the case where the turbines operate at their capacity, a similar expression can be derived.

The application to the 72 GW scenario is briefly illustrated here: With $H$ = 700 m, $W = L \approx$ 85 090 m, $C_d$ = 0.001, $\eta$ = 0.42, and $N$ = 6033, this results in a factor of $f_{red}$ = 870 m/ (870 m + 1404 m) = 0.38. When yields depends on wind speed, this factor implies that wind extraction causes the yields to drop to 38%, a 62% reduction, while wind speeds have only dropped by 28%. However, this is only a partial aspect of the overall yield, as there are still times when the turbines operate at their capacity. Therefore, the reduction in Figure 3 is less dramatic at 40%. The various components of the kinetic energy balance can then be determined by combining the observed energy flux density (505 W m$^{-2}$ in the median at $v_{in}$ = 9.4 m/s) with the parameters and equations.

These KEBA calculations are available as a spreadsheet for yield estimates on the Internet [8].

**FIG. 5 KINETIC ENERGY COMPONENTS**

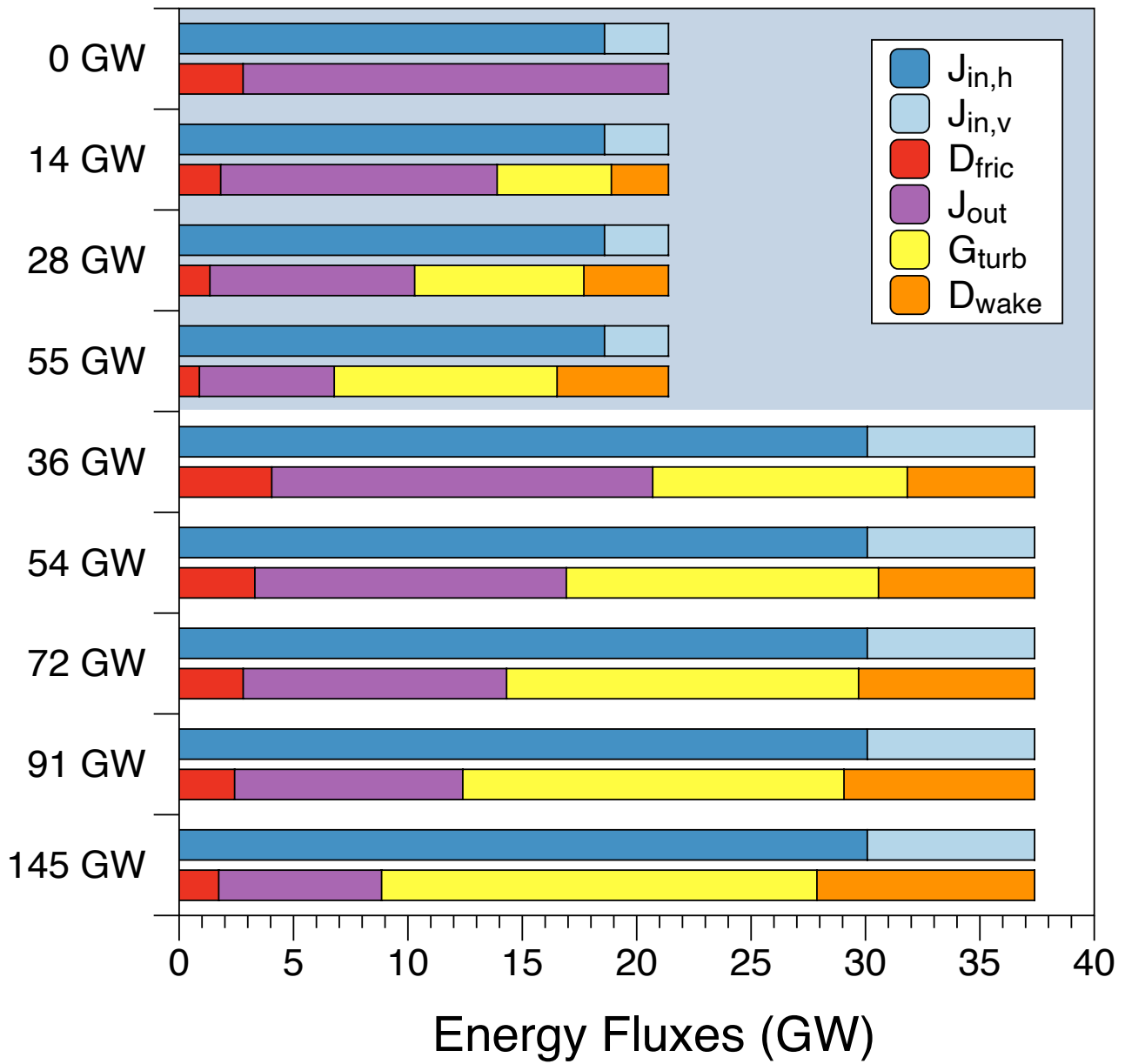

Components of the kinetic energy balance for the scenarios of Figure 3, the upper light blue section again applies to Area 1 alone, the lower white section to Area 1 and 2 together. The values are estimated with the KEBA approach (colouring as in Figure 4, explanation of the symbols in the legend and box "KEBA: Kinetic energy balance of the atmosphere").

**FIG. 6 WIND SPEEDS**

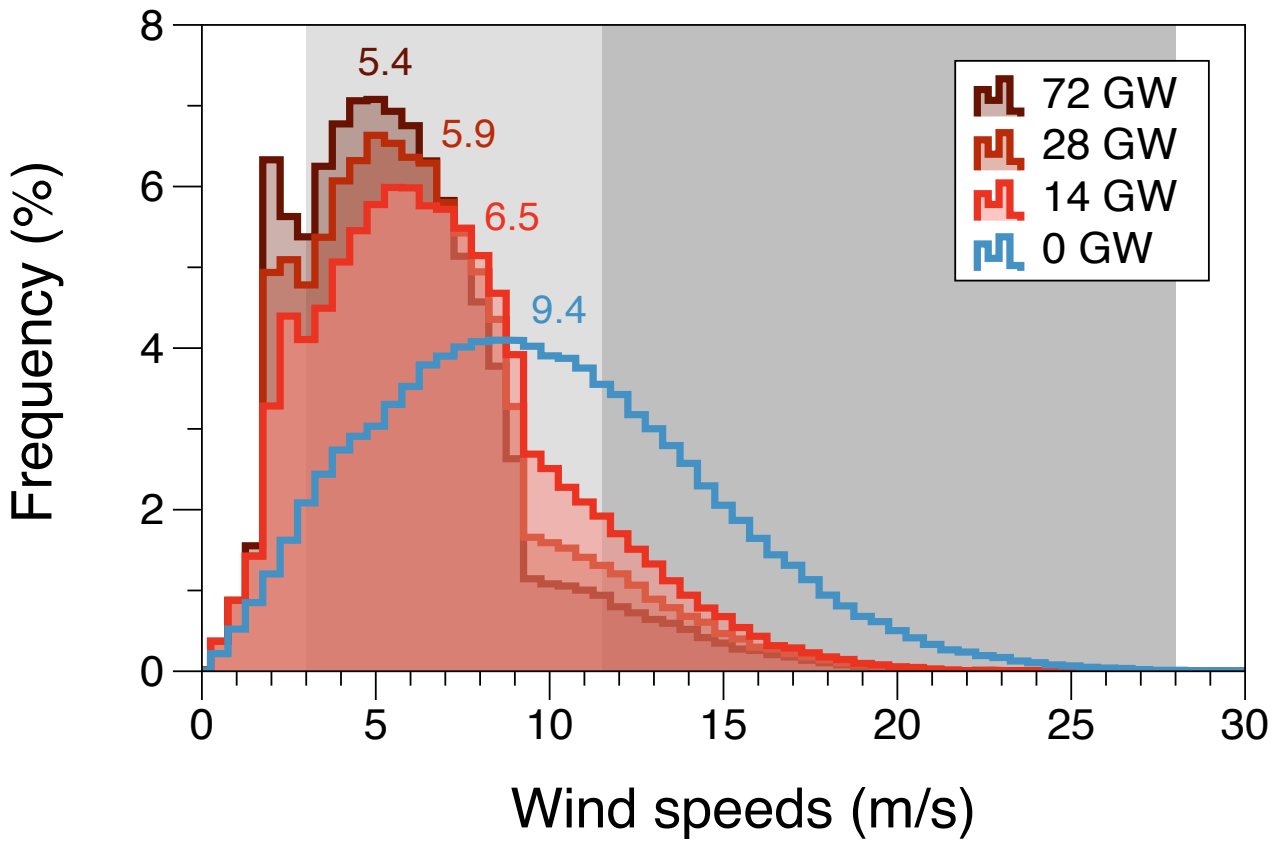

*Frequency distribution of wind speeds of three scenarios illustrating the shift to lower wind speeds with more wind energy use. The numerical values indicate the median of the respective distributions.*

## Conclusions

Overall, this gives us a differentiated picture of the contribution that offshore wind energy can make to the energy transition: On the one hand, the potential to generate electricity is huge, even with the associated, significant reductions due to wind extraction from the turbines. For example, the 72 GW scenario can cover more than a third of Germany's current electricity consumption. On the other hand, the use is much more efficient if wind farms are less dense and distributed over larger areas. This can be seen in a direct comparison of the scenarios 55 GW installed in Area 1 and 54 GW installed in both areas (Figures 3 and 5). In the latter case, the reduction effect is much smaller, as the turbines are distributed over a much larger area.

This weakening effect will therefore play an increasingly important role in the expansion of offshore wind energy. It is independent of the technology, the size of the turbines or the positioning of the turbines in the wind farm. After all, the main effect has to do with what the turbines are there to do: To extract energy from the wind in order to generate electricity with it.

## Summary

*A significant contribution to the energy transition is expected from offshore wind energy in the German Bight. Due to the strong and steady winds, offshore electricity generation appears to be very efficient. For 2050, the German government assumes an installed capacity of 70 gigawatts, a tenfold increase compared to today. But what happens when so many wind turbines draw their energy from the wind? This can be easily determined with the help of the kinetic energy balance of the atmosphere above the wind farms. Since the input of kinetic energy is limited, the more wind energy is used, the lower the wind speeds in the region must be, and with them the efficiency of the turbines. So less electricity is generated than would be expected without this effect. At 70 GW, that would reduce electricity generation by as much as 40%. Still, it could meet a large part of the current electricity demand. For the efficient use of wind energy at sea, it is therefore advisable to plan wind farms as widely dispersed as possible in order to reduce their influence on the wind fields.*

## The author

Axel Kleidon studied physics and meteorology at the University of Hamburg and Purdue University, Indiana, USA. He received his doctorate from the Max Planck Institute for Meteorology in 1998 for his work on the influence of deep-rooted vegetation on the climate system. He subsequently conducted research at Stanford University in California and at the University of Maryland. Since 2006, he has headed the independent research group "Biospheric Theory and Modelling" at the Max Planck Institute for Biogeochemistry in Jena. His research interests range from the thermodynamics of the Earth system to the natural limits of renewable energy sources.

**Address**

Dr. Axel Kleidon, Max Planck Institute for Biogeochemistry, Postfach 10 01 64, 07701 Jena. axel.kleidon@bgc-jena.mpg.de